\documentclass[twocolumn,amsmath,amssymb,10pt,superscriptaddress,a4paper,letterpaper,fleqn,floatfix,nofootinbib,prc,aps]{revtex4-1}
\usepackage{amssymb}
\usepackage{graphicx}
\usepackage[breaklinks=true,linkbordercolor={1 1 1}]{hyperref}
\usepackage{color}

\def\Gammabar{\overline{\Gamma}}
\newcommand{\tiltfrac}[2]{\,^#1\!/_#2}

\begin{document}
%%% **********************************************************************

\title{Impact of alternative transmission coefficient parameterizations on Hauser-Feshbach theory}

\author{D.A. Brown}
\email{dbrown@bnl.gov}
\affiliation{National Nuclear Data Center, Brookhaven National Laboratory, Upton, NY 11973, USA}
\author{G.P.A. Nobre}
\email{gnobre@bnl.gov}
\affiliation{National Nuclear Data Center, Brookhaven National Laboratory, Upton, NY 11973, USA}
\author{M.W. Herman}
%\email{mwherman@bnl.gov}
\affiliation{National Nuclear Data Center, Brookhaven National Laboratory, Upton, NY 11973, USA}
%\affiliation{Los Alamos National Laboratory, Los Alamos, NM}

\date{\today} 

\begin{abstract}
{
We investigate different formulations of the transmission coefficient $T_c$, including the form implied by Moldauer's ``sum rule for resonance reactions'' [P.A. Moldauer, Phys. Rev. Lett. {\bf 19}, 1047 (1967)], the SPRT method [G. Noguere, {\em et al.} EPJ Web Conf. 146, 02036 (2017)] and the Moldauer-Simonius form [M. Simonius, Phys. Lett. {\bf 52B}, 279 (1974); P.A. Moldauer, Phys. Rev. {\bf 157}, 907 (1967)].  Within these different formulations, we compute the neutron transmission coefficients in the resolved and unresolved resonance regions, allowing a direct comparison with the transmission coefficients computed using an optical model potential.  For nuclei for which there are no measured resonances, these approaches allow one to predict the average neutron resonance parameters directly from the optical model and level densities.  Some of the approaches are valid in both the strong and weak coupling limits (i.e., any value of the average width and mean level spacing).  Finally, both the Moldauer-Simonius and Moldauer's Sum Rule forms approaches suggest that superradiance, that is, the quantum chaotic enhancement of certain channels, may be a common phenomena in nuclear collisions.  Our results suggest why superradiance has been previously overlooked.  We apply our approach to neutron reactions on the closed shell $^{90}$Zr nucleus and the mid-shell $^{197}$Au nucleus.
}
\end{abstract}
\maketitle

% ===========================================================================
\section{Introduction}
% ===========================================================================
\label{sec:intro}

For neutron-induced reactions below 20 MeV incident energy, the unresolved resonance region (URR) connects the fast neutron range with the resolved resonance region (RRR).  The URR is problematic since the resonances in this region are not resolvable experimentally, yet the fluctuations in the neutron cross sections play a discernible and technologically important role: the URR in a typical nucleus is in the 100 keV -- 2 MeV window, where fission spectra peak.  The URR also represents the transition between theoretical approaches.  In the RRR, R-matrix theory is used to describe the shape and correlations between resolved resonances in angle-differential cross sections.  In the fast region, Hauser-Feshbach theory, with the Width Fluctuation Correction (WFC), is used to accurately describe the cross sections.  

Given our lack of knowledge of resonance positions and widths in the URR, we can only determine the average resonance spacing $D$, the average channel widths $\overline{\Gamma}_c$ and the number of degrees of freedom $\nu_c$ for each channel.  Here a channel denotes the two incoming/outgoing particles and all the quantum numbers needed to specify their state (for our purposes only the orbital angular momentum $L$ and total angular momentum $J$).  As the cross section fluctuates strongly in the URR, at best we can describe the probability distribution of the cross section in terms of $D, \overline{\Gamma}_c$ and $\nu_c$.  As a first step toward determining the full probability distribution,  we focus on the energy average cross sections in the URR.  The Gaussian Orthogonal Ensemble (GOE) triple integral result of Verbaarschot, Weidenm\"uller, and Zirnbauer~\cite{VZW} is believed to provide an exact solution for the energy averaged cross section.  Unfortunately, this result is both difficult to interpret physically and numerically expensive to use in practice so it is not appropriate for gaining insight into the physics of compound nuclear reactions.  The Hauser-Feshbach equation with Moldauer's Width Fluctuation Correction (WFC)~\cite{Moldauer1975a} is known to be both easier to use and simpler to interpret~\cite{Kawano2015}.  Moldauer's WFC was derived in the weak coupling limit ($x_c=\pi \overline{\Gamma}_c/D \ll 1$), but it is regularly used outside its region of validity.

To move beyond the weak coupling limit, it is necessary to understand the connection between the transmission coefficients $T_c$ used at higher energies and  $D, \overline{\Gamma}_c$ and $\nu_c$ used in the URR.  In this paper, we will investigate three parameterizations of $T_c$, the SPRT method~\cite{Noguere2016}, Moldauer's ``optical model'' form which we call the Moldauer-Simonius form~\cite{Simonius1974,Moldauer1967a}, and the result that is implied by Moldauer's ``sum rule for resonance reactions''~\cite{{Moldauer1967}}.  We will also investigate the weak coupling limits of all three forms.  

Any prescription that connects the average resonance widths $\bar{\Gamma}_c$ and level spacings $D$ to transmission coefficients $T_c$ allows for a unified framework that connects the RRR, URR and fast regions.  As the optical model provides a predictive theory for the transmission coefficients one could in principle extract the average neutron resonance width for nuclei off the valley of stability using an extrapolated level density.   That said, both of Moldauer's results appear to be have been ignored because both predict a singularity in the compound nuclear cross section as one approaches the strong coupling limit $\bar{\Gamma}_c/D \ge 1$.  Below we argue that this singularity is not reached in practice.  More interestingly, this singularity appears to be a manifestation of superradiance~\cite{Auerbach2011}, namely, the quantum chaotic enhancement of certain channels.  We present reasons why superradiance may have been overlooked in the past.

We are not the first ones to consider using Hauser-Feshbach theory with the WFC to characterize the average cross section in the URR.  This is the basis of the $^{238}$U evaluation of Fr\"ohner~\cite{Froehner1989,Froehner1992} and the later evaluations of Sirakov {\em et al.}~\cite{Capote}.  It is also the basis for the SPRT method in Ref.~\cite{Rich2009}, the MC$^2$-II method described in the Appendix D.2 of the ENDF Formats Manual~\cite{ENDFFormat}, and built into the SESH evaluation code by Fr\"ohner~\cite{Froehner1968}.  In all these approaches, the weak coupling limit form of the transmission coefficients is implicitly assumed.  %Our observations provide clues how to modify evaluation procedures to not only connect the different physical regimes, but to provide coherent covariance estimates over the entire energy range of practical importance.

The outline of this paper is as follows.  In Section \ref{sec:background}, we begin by introducing compound nuclear reaction cross sections, laying the ground work for the the transmission coefficient.  In Section \ref{sec:Tc}, we describe the different transmission coefficient formulations.  We continue by exploring the transmission coefficients in Section \ref{subsec:compareTc} and their impact on the compound nuclear cross sections in Section \ref{sec:superradiance}.  We apply these results to $^{90}$Zr and $^{197}$Au in Sections \ref{sec:application90Zr} and \ref{sec:application197Au} respectively and find that we can describe the neutron transmission coefficients at all relevant energies.  We also compute the compound reaction cross sections and demonstrate that the superradiant form of the cross section does not introduce any unwanted changes to the calculated cross sections.  In Section \ref{sec:conclusion} we conclude with a brief statement of the implications of our observations.

% ===========================================================================
\section{Theoretical Background}
% ===========================================================================
\label{sec:background}

The $(n,x)$ reaction cross section is found in many sources and textbooks, such as Refs.~\cite{Froehner,FroebrichLipperheide}, and is given as
\begin{equation}
	\sigma_{(n,x)}
		=\sum_{J^\Pi}\sum_{\substack{a\in n,J^\Pi\\ b\in x,J^\Pi}} \sigma_{ab}.
	\label{xs:reaction}
\end{equation}
Here we have decomposed the terms by total angular momentum and parity ($J^\Pi$) and by the remaining quantum numbers grouped together collectively as a channel index (denoted by the letters $a, b, c...$), including the orbital angular momentum $L$.  The per-channel reaction cross section can be written in terms of the scattering matrix ${\bf S}$ as
\begin{equation}
	\sigma_{ab}=\frac{\pi}{k^2_a}g_a\left|\delta_{ab}-S_{ab}\right|^2.
	\label{xs:channelReaction}
\end{equation} 
The statistical factor $g_c=(2J_c+1)/\left[(i_c+1)(2I_c+1)\right]$ and $i_c$ and $I_c$ are the intrinsic spins of the projectile and target respectively.  With this, the total cross section for incident neutrons is 
\begin{equation}
	\sigma_{(n,\textrm{tot})}
		=\sum_{(n,x)} \sigma_{(n,x)}
		\equiv \sum_{J^\Pi}\sum_{a\in n,J^\Pi} \sigma_{a}.
		\label{xs:total}
\end{equation}
Here the total channel cross section $\sigma_{a}$ can be written in terms of the scattering matrix ${\bf S}$ as 
\begin{equation}
	\sigma_c=\frac{2\pi}{k_c^2}g_c\left\{1-\Re S_{cc}\right\}.
\label{eq:totalChannelCrossSection}
\end{equation}
One can also compute the angle differential cross section for two body final state reactions using the Blatt-Biedenharn formalism \cite{Froehner,BlattBeidenharn}, but for simplicity we will not do that here.

We now write the energy averaged S-matrix as $S_{cc'}=\left<S_{cc'}\right>+S_{cc'}^{\textrm{fl}}$, i.e. as the sum of a smooth background scattering matrix $\left<S_{cc'}\right>$ and a fluctuating term $S_{cc'}^{\textrm{fl}}$ with $\left<S_{cc'}^{\textrm{fl}}\right>=0$.  The energy-averaged $\left<f\right>$ of a function $f(E)$ is the usual
\begin{equation}
	\begin{split}
	\left<f\right>&\equiv\frac{\Delta E}{\pi}\int^{\infty}_{-\infty}  dE'\frac{f(E')}{(E-E')^2+(\Delta E)^2}
	\label{eq:fancyenergyaverage}\\
	              &=f(E+i\Delta E).
	\end{split}
\end{equation}
This energy averaging scheme is only appropriate if $f$ has no poles with $\Im E > 0$ 
and $f$ is bounded as a function of energy as $|E|\rightarrow\infty$.   With this, Eq. \eqref{eq:fancyenergyaverage} can be performed as a complex contour integration where the contour is a semi-circle in the upper half-plane that only surrounds the pole at $E+i\Delta E$.  

Separating the scattering matrix into smooth and fluctuating parts allows us to write the total channel cross section as 
%$\left<\sigma_{\textrm{tot}}\right>$ using Eq. \eqref{xs:total}, $\left<\sigma_{(n,x)}\right>$ using Eq. \eqref{xs:reaction}, 
\begin{equation}
	\left<\sigma_c\right>=\frac{2\pi}{k_c^2}g_c\left\{1-\Re \left<S_{cc}\right>\right\},
	\label{eq:aveTotalPerChannel}
\end{equation}
and the channel-channel cross section as
\begin{equation}
	\left<\sigma_{ab}\right>=\frac{\pi}{k^2_a}g_a\left|\delta_{ab}-\left<S_{ab}\right>\right|^2 + \frac{\pi}{k^2_a}g_a\left<\left|S_{ab}^{\textrm{fl}}\right|^2\right>.
	\label{eq:aveChannelReaction}
\end{equation}
The first term on the right of Eq. \eqref{eq:aveChannelReaction} is conventionally defined as the direct cross section $\sigma_{ab}^{\textrm{dir}}$ and the second term as the compound nuclear cross section $\sigma_{ab}^{\textrm{cn}}$. 

Up to a normalization factor, the compound nuclear term is usually written as the well known Hauser-Feshbach formula with the Width Fluctuation Correction~\cite{Moldauer1975a,Moldauer1975b}
\begin{equation}
	\left<\left|S_{ab}^{\textrm{fl}}\right|^2\right>=\left<\left|S_{ab}\right|^2\right>-\left|\left<S_{ab}\right>\right|^2\approx \frac{1}{D}\frac{\Gammabar_a\Gammabar_{b}}{\sum_{c}\Gammabar_{c}}{\cal W}_{ab}(\vec{\Gamma}),\label{eq:HauserFeshbach}
\end{equation}
where ${\cal W}_{ab}(\vec{\Gamma})$ is the Width Fluctuation Correction (WFC) and is a function of the average widths of all relevant channels.  For economy of notation, we have condensed this dependence into a vector of widths.  
Defining the absorption cross section for channel $a$ as $\sigma^{\textrm{abs}}_a~=~\tiltfrac{{\pi~g_a\Gammabar_a}}{{Dk^2_a}}$, we write
\begin{equation}
	\sigma_{ab}^{\textrm{cn}}=\sigma^{\textrm{abs}}_a\frac{\Gammabar_{b}}{\sum_{c}\Gammabar_{c}}{\cal W}_{ab}(\vec{\Gamma}).
	\label{eq:HF}
\end{equation}

The WFC was originally derived by Dresner \cite{Dresner1957} and by Lane and Lynn \cite{Lane1957} under the assumption that resonances are widely spaced so interference effects between resonances can be ignored, essentially making the mean level spacing $D$ much larger than the average widths $\bar{\Gamma}$.  The cross sections then simplify to a form equivalent to the single level Breit Wigner approximation~\cite{Lane1957,Froehner}.  Under these conditions, one assumes that the resonance widths $\Gamma_c$ follow a $\chi^2$ distribution with $\nu_c$ degrees of freedom, arriving at
\begin{equation}
\begin{split}
	\lefteqn{{\cal W}_{ab}(\vec{\Gamma})=
		\left(1+\delta_{ab}\frac{2}{\nu_a}\right)}&\\
		&\times\int_0^\infty dx \prod_c\left( 1+\frac{2\Gammabar_c}{\nu_c\sum_i\Gammabar_i}x\right)^{-\delta_{ac}-\delta_{bc}-\nu_c/2}.
		\label{eq:WFC}
\end{split}
\end{equation}
%This derivation is elaborated on in Appendix B along with extensions to the full cross section covariance.  
Improvements to Dresner's and Lane and Lynn's original result, such as Moldauer's approach \cite{Moldauer1975a}, are based on phenomenological fits of transmission coefficient dependent $\nu_c(T_c)$ using numerical simulations.
% data or results of the GOE Triple Integral.   These fits basically extend the validity of Eq. \eqref{eq:WFC} to the strong coupling limit.  Modern WFC are believed to be robust if one uses the most recent parameter choices e.g. those determined by Kawano and Talou \cite{Kawano2015}

%If one includes the Dresner simplification for gamma channels in the limit have many gamma channels (doesn't work well with light nuclei or magic nuclei), define $\Gammabar_\gamma=\sum_{c\in\gamma}\Gammabar_c$ and $\Gammabar=\sum_c\Gammabar_c$:
%\begin{equation}
%\begin{split}
%\prod_{c\in\gamma}\left(1+\frac{2\Gammabar_c}{\nu_c \Gammabar}x\right)^{-\nu_c/2}
%	&\approx\lim_{\overline{\nu}_\gamma\rightarrow\infty}\left(1+\frac{2\Gammabar_\gamma}{\overline{\nu}_\gamma \Gammabar}x\right)^{-\overline{\nu}_\gamma/2}\\
%	&=e^{-x\Gammabar_\gamma/\Gammabar}.
%\end{split}
%\end{equation}

The Hauser-Feshbach formula with the WFC as given in most textbooks \cite{Satchler,FroebrichLipperheide} is written in terms of the transmission coefficient $T_c=1-\left|\left<S_{cc}\right>\right|^2$.  Noting that in the weak coupling limit $T_c\approx 2\pi\overline{\Gamma}_c/D$, one usually replaces 
\begin{equation}
	\frac{\Gammabar_{b}}{\sum_{c}\Gammabar_{c}} \rightarrow \frac{T_{b}}{\sum_{c}T_{c}}
	\label{eq:replacementOld}
\end{equation}
in Eq. \eqref{eq:HF} and Eq. \eqref{eq:WFC}, giving the more traditional 
\begin{equation}
	\sigma_{ab}^{\textrm{cn}}\approx\sigma^{\textrm{abs}}_a\frac{T_{b}}{\sum_{c}T_{c}}{\cal W}_{ab}(\vec{T}),
	\label{eq:HF_Tc}
\end{equation}
where $\sigma^{\textrm{abs}}_a\approx\tiltfrac{{g_a T_a}}{{2k^2_a}}$.  In the Section \ref{sec:superradiance}, we will argue that this conventional form is incomplete and must be modified, giving rise to a form that predicts superradiance.

% ===========================================================================
\section{Transmission coefficients}
% ===========================================================================
\label{sec:Tc}

We now investigate alternate formulations of the transmission coefficients and explore their implications as they are the key to relating the transmission coefficient to the average widths and level spacings.  We will explore three separate transmission coefficient models and their weak coupling limits.

\subsection{The SPRT method}
% --------------------------
\label{subsec:SPRT}

Our first approach is based on the SPRT method, which we now outline.  The SPRT method has been championed by Noguere \textit{et al.} \cite{Noguere2016,Rich2009} precisely because it provides a direct connection between R-matrix parameters and the transmission coefficient. As the method is based on work from Moldauer \cite{Moldauer1963}, we outline his arguments.  

It is clear from Eqs. \eqref{eq:aveTotalPerChannel} and \eqref{eq:aveChannelReaction} that we must consider the energy averaged S-matrix, even though it does not appear directly in the final Hauser-Feshbach equation with WFC.  The S-matrix can be written in terms of the R-matrix through
\begin{equation}
	\begin{split}
	S_{cc'}&=e^{-i(\phi_c+\phi_{c'})}
			P_c^{1/2}\\
	        &\times\left\{ \left[{\bf 1}-{\bf R}{\bf L}^0\right]^{-1} \left[{\bf 1}-{\bf R}{\bf L}^{0*}\right] \right\}_{cc'}
	        P_{c'}^{-1/2}.
	\end{split}
	\label{eq:SthroughR}
\end{equation}
Here the components of the R-matrix are given by
\begin{equation}
	R_{cc'}=\sum_\lambda\frac{\gamma_{\lambda c}\gamma_{\lambda c'}}{E_\lambda-E}.
\end{equation}
Where $\gamma_{\lambda c}=\sqrt{2P_c\Gamma_{\lambda c}}$ are the reduced widths and $E_\lambda$ are the pole energies.  The other parameters in Eq. \eqref{eq:SthroughR} are the hard-sphere phase-shift $\phi_c$ and the hard-sphere penetrability $P_c$.  The matrix ${\bf L}^0$ is related to the penetrability, the hard-sphere shift factor $S_c$, and the R-matrix boundary parameters $B_c$ through
\begin{equation}
	L^0_{cc'}\equiv(S_c+iP_c-B_c)\delta_{cc'}.
\end{equation}

Moldauer replaces the R-matrix with the energy averaged R-matrix on an interval $\Delta E$,
\begin{equation}
	\begin{split}
	\left<S_{cc'}\right>&\approx e^{-i(\phi_c+\phi_{c'})}
			P_c^{1/2}\\
	        &\times\left\{ \left[{\bf 1}-\left<{\bf R}\right>{\bf L}^0\right]^{-1} \left[{\bf 1}-\left<{\bf R}\right>{\bf L}^{0*}\right] \right\}_{cc'}
	        P_{c'}^{-1/2}.
	\end{split}
	\label{eq:SthroughAveR}
\end{equation}
Moldauer then shows that the energy averaged R-matrix is related to the pole strength $s_c$ and the distant level parameter $R^\infty_c$ through
\begin{equation}
	\left<R\right>_{cc}=i\pi s_c+R^\infty_c.
\end{equation}
The distant level parameter is the Hilbert transform of the pole strength
\begin{equation}
	R^\infty_c=PV\int_{-\infty}^{\infty}dE'\frac{s_c(E')}{E'-E}.
\end{equation}
and the pole strength is
\begin{equation}
	s_c=\frac{\sum_\lambda \gamma_{\lambda c}^2}{\Delta E},
	\label{eq:scSum}
\end{equation}
on the averaging interval $\Delta E$.  The sum in Eq. \eqref{eq:scSum} can be approximated as an ensemble average as described in the Appendix.  Using Eq.~\eqref{aeq:4} from the Appendix, 
\begin{equation}
	s_c\approx\frac{\Gammabar_{c}}{2DP_c}.\label{eq:sc}
\end{equation}
%The distant level parameter $R^\infty_c$ through
%\begin{equation}
%\left<R\right>_{cc}=i\pi\frac{\left<\left<\gamma_c^2\right>\right>}{D}+R^\infty_c
%\end{equation}

With these, and the conventional definition of the transmission coefficient $T_c=1-|\left<S_c\right>|^2$, we have
\begin{equation}
	T_c^{\textrm{SPRT}}=\frac{2x_c}{(1+x_c/2)^2+(P_c{R^\infty_c})^2}.
	\label{eq:TcSPRT}
\end{equation}
Here we define Verbaarschot, Weidenm\"uller, and Zirnbauer's parameter $x_c$ as $x_c\equiv\pi\Gammabar_c/D=2\pi P_c s_c$.  Noguerre \emph{et al.} provide a modified form of this expression that includes some direct reaction effects, amounting to adding an additional term to $T_c$.

The transmission coefficient implied by Eq. \eqref{eq:TcSPRT} rises from zero at $x_c~\rightarrow~0$ to its peak value of 
\begin{equation*}
Tc(x_c=2)=(1+(P_cR_c^\infty/2)^2)^{-1}
\end{equation*}
 %at $x_c=2$ 
 and then decreases monotonically as $x_c~\rightarrow~\infty$.  The peak value of the transmission coefficient only reaches unity when the distant level parameter is ignored.  Moldauer argues that the peak occurs at the so-called optical model peak in the total cross section (typically around a few MeV).  We disagree with this assertion as the peak in the total cross section is controlled by the energy dependence of the hard sphere phase $\phi_c$ as one can see by combining Eqs.~\eqref{eq:SthroughR} and~\eqref{eq:totalChannelCrossSection}.

\subsection{Moldauer's Sum Rule}
% ------------------------------
\label{subsec:SR}

Moldauer provided an alternative approach to computing the average scattering matrix in Ref. \cite{Moldauer1967}, leading to his ``sum rule of the resonance region'', hereafter called the Moldauer's Sum Rule or just Sum Rule.  Following Moldauer \cite{Moldauer1978,Moldauer1967}, we write the scattering matrix in the resonance pole form as
\begin{equation}
	S_{cc'}=e^{-i(\phi_c+\phi_{c'})}\left(S^0_{cc'}+i\sum_\lambda\frac{G_{\lambda cc'}}{E-{\cal E}_\lambda}\right). \label{eq:fullSMatrix}
\end{equation}
Here the width matrix $G_{\lambda cc'}$ and pole energy ${\cal E}_\lambda$, are easiest understood in the multilevel Breit-Wigner (MLBW) approximation of the R-matrix \cite{Froehner}: 
\begin{equation}
	G_{\lambda cc'}=\Gamma_{\lambda c}^{1/2}\Gamma_{\lambda c'}^{1/2}
\end{equation}
and 
\begin{equation}
	{\cal E}_\lambda=E_\lambda+\Delta_\lambda-i\Gamma_\lambda/2.
\end{equation}
Similar identifications can be made for other formulations of the R-matrix.  Here $S^0_{cc'}$ is a smooth background part of the S-matrix which is usually taken as the identity matrix.  We comment that the poles at ${\cal E}_\lambda$ have $\Im {\cal E}_\lambda <0$ due to causality.  Any unitary transform of the $G_{\lambda cc'}$ and $S^0_{cc'}$ matrices cannot change the pole structure.  %Therefore results are valid with or without the Englebrecht-Weidenm\"uller transformation.

\begin{figure}
	\includegraphics[width=0.48\textwidth]{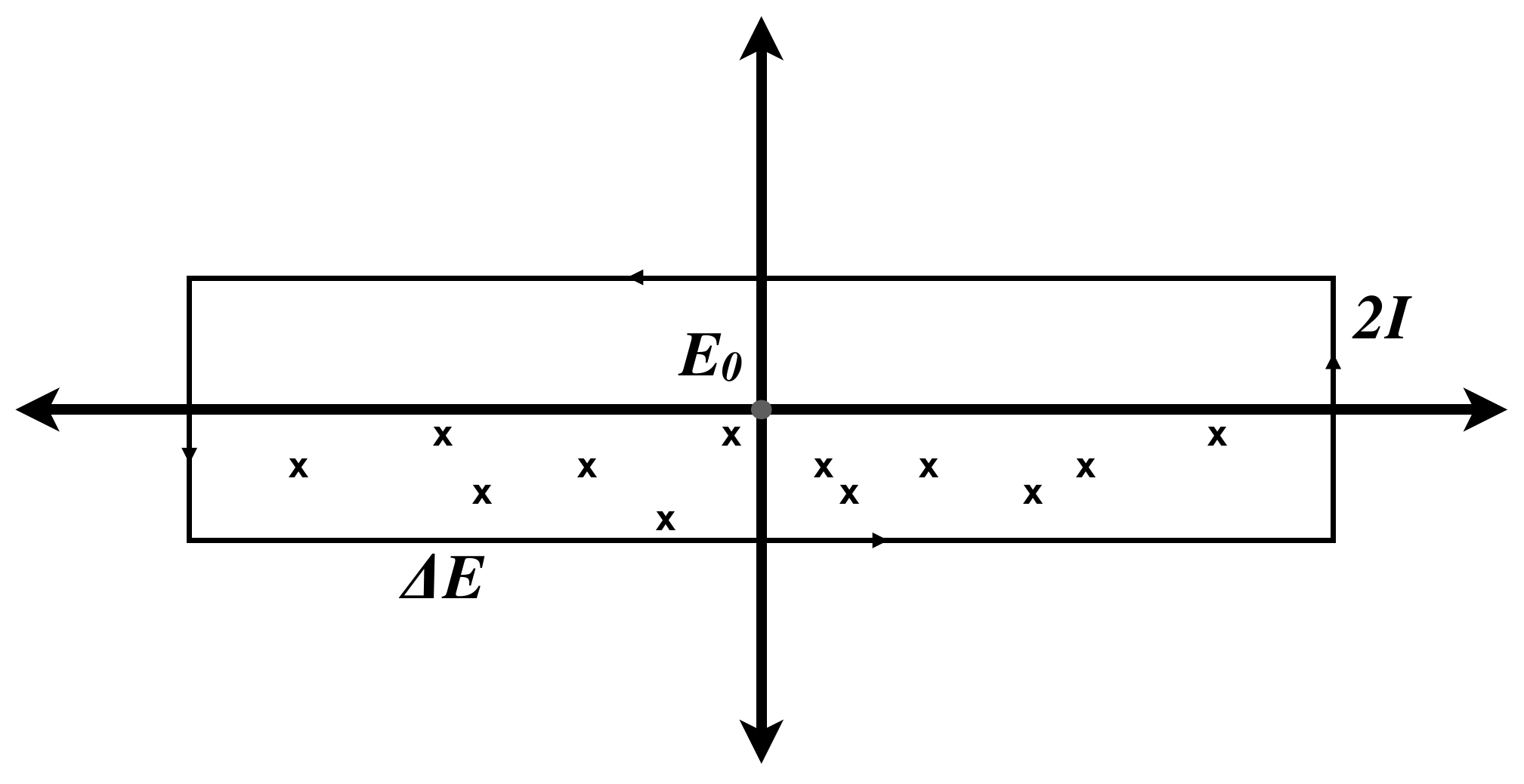}
	\caption{\label{fig:contour}Contour in the complex energy plane used in the derivation of Moldauer's sum rule.  The contour is a counter clockwise rectangle of length along the real axis $\Delta E$ and width in the imaginary direction of $2I$.  The scattering matrix poles are all in the lower half-plane due to causality.}
\end{figure}

Rather than use the Lorentian shaped smoothing from Eq. \eqref{eq:fancyenergyaverage}, 
%Moldauer considered the contour shown in Fig. \ref{fig:contour} for his energy averaging.  
Moldauer considered a different contour: a counter clockwise rectangle of length $\Delta E$ along the real axis and width in the imaginary direction of $2I$, shown in Fig. \ref{fig:contour}.
With this, he found 
\begin{eqnarray*}
	\oint dE S_{cc'}(E) &=& 2\pi\sum_\lambda G_{\lambda cc'}\\
 					&=& \Delta + \left(S_{cc'}(E_0-iI)\right.\\
					&& \left.-S_{cc'}(E_0+iI)\right)\Delta E
\end{eqnarray*}
where $\Delta$ is the contribution from the ends at $E\rightarrow\pm\infty$.  Using Eq. \eqref{eq:fancyenergyaverage}, unitarity, and analyticity, we have
\begin{equation}
	2\pi\sum_\lambda G_{\lambda cc'} - \Delta = \left( \left<S_{cc'}^*\right>^{-1}-\left<S_{cc'}\right>\right) \Delta E.
	\label{eq:sumrule}
\end{equation}

Moldauer then argues that if we perform an ensemble average over sets of poles, $\Delta$ will vanish\footnote{We suspect, without substantial proof, that $\Delta$ is related to the distant level parameter $R^\infty_c$.} and we may approximate 
\begin{equation}
	\sum_\lambda G_{\lambda cc'}\approx\frac{\Delta E}{D}\left<\left<G_{cc'}\right>\right>.
	\label{eq:Gthingee}
\end{equation}
Here $D$ is the average resonance pole spacing and
$\left<\left<f(X)\right>\right>$ is an ensemble average of the quantity in the double brackets.  In other words, $\left<\left<f(X)\right>\right>=\int dX{\cal P}(X)f(X)$ where ${\cal P}(X)$ is the probability distribution function for quantity $X$.  Eqs. \eqref{eq:sumrule} and \eqref{eq:Gthingee} constitute Moldauer's Sum Rule.

%\textcolor{red}{Assuming the ensemble average is over Porter-Thomas distributions of resonance widths and approximately equally spaced energies, we arrive at $\left<\left<G_{cc'}\right>\right>=\Gammabar_c^{1/2}\Gammabar_{c'}^{1/2}$. % and $\tau_c=\pi\Gammabar_c/D$.
%Moldauer points out that 
%\begin{equation}
%\left|\left<\left<G_{cc}\right>\right>\right|\le\left<\left<|G_{cc}|\right>\right>=\left<\left<|g_{\mu c}|^2\right>\right>.
%\end{equation}
%We essentially use the upper bound of this inequality.  Moldauer argues that this limit is saturated in his simulations \cite{Moldauer1967a}.} 
%\textcolor{red}{When does $\left<\left<|g_{\mu c}|^2\right>\right>=\left<\left<|\gamma_{\mu c}|^2\right>\right>$?  Or better yet, when does $G_{\lambda cc'}=\gamma_{\mu c}\gamma_{\mu c'}$?}

In the presence of direct reactions, the average S-matrix is not diagonal.  Although the $S$-matrix is unitary, the energy averaged $S$-matrix is not and $\left<\mathbf{S}\right>$ and $\left<\mathbf{S}^*\right>$ cannot simultaneously be diagonalized.  Englebrecht and Weidenm\"uller's solution was to diagonalize Satchler's transmission matrix \cite{Engelbrecht1973}:
\begin{equation}
	\rho_{ab}=1-\sum_c\left<S_{ac}\right>\left<S^*_{bc}\right>=\sum_cU^\dagger_{ac}\varrho_cU_{cb}.
\end{equation}
Here $\mathbf{U}$ is the unitary matrix that diagonalizes the transmission matrix with eigenvalues $\varrho_c$.  
The eigenvalues of the transmission matrix $\varrho_a$ play the role of transmission coefficients in Englebrecht and Weidenm\"uller's reformulation of Hauser-Feshbach theory.  We note that the transmission coefficient is usually identified as the trace of the transmission matrix:
\begin{equation}
	T_a=\rho_{aa}=1-\sum_{c}|\left<S_{ac}\right>|^2=\sum_c\left|U_{ac}\right|^2\varrho_c,
\end{equation}
so in the absence of direct reactions, 
\begin{equation}
	T_a=\varrho_a= 1-|\left<S_{aa}\right>|^2.
\end{equation}

Englebrecht and Weidenm\"uller continue to show that, with the aid of Moldauer's Sum Rule,
\begin{equation}
	\frac{2\pi}{D}\sum_{cc'}U_{ac}\left<\left<G_{cc'}\right>\right>U^*_{c'b} =\delta_{ab} \varrho_a (1-\varrho_a)^{-1/2}e^{2i\phi_a}.
\end{equation}
We identify $x_a$ as 
\begin{equation}
	x_a = \frac{\pi}{D}\left| \sum_{cc'}U_{ac}\left<\left<G_{cc'}\right>\right>U^*_{c'a} \right|.
	\label{eq:x}
\end{equation}
With this, we find 
\begin{equation}
	\varrho_a=2x_a\left[\sqrt{x_a^2+1}-x_a\right].
	\label{eq:sumrulerhoc}
\end{equation}

If we may ignore direct reactions, then both the energy averaged S-matrix and $\left<\left<G_{cc'}\right>\right>$ are diagonal.  Furthermore, we have $\left<\left<G_{cc'}\right>\right>\approx\delta_{cc'}\left<\left<\Gamma_c\right>\right>$ in the MLBW approximation so Eq. \eqref{eq:x} gives $x_a=\pi\Gammabar_{a}/D$ (hence our identification of $x_a$).  This gives us Moldauer's Sum Rule result for the transmission coefficient
\begin{equation}
	T_a^{\textrm{SR}} = 2x_a\left[\sqrt{x_a^2+1}-x_a\right].
	\label{eq:sumruleTc}
\end{equation}

In the presence of direct reactions, the MLBW approximation give us 
$\left<\left<G_{cc'}\right>\right>=\left<\left<\Gamma_{c}^{1/2}\Gamma_{c'}^{1/2}\right>\right>$.  Inserting Eq. \eqref{eq:Gthingee} into Eq. \eqref{eq:x}, we see that the Englebrecht--Weidenm\"uller transform correlates the different channel widths of a given resonance.
Given this, it is difficult to make progress from this point as we do not know the joint probability distribution for the widths for different channels of the same resonance.    In the future we hope to use this fact to generalize Eq. \eqref{eq:sumruleTc} for cases where direct reactions cannot be neglected.

\subsection{The Moldauer--Simonius form}
% --------------------------------------
\label{subsec:MS}

In 1967, Moldauer published the results of a series of numerical and analytic studies motivating the following form of the transmission coefficient \cite{Moldauer1967a}:
\begin{equation}
	T_c^\textrm{MS}=1-\exp{(-2x_c)}.
	\label{eq:opticalModelTc}
\end{equation} 
He then conjectured that this form is general and indeed it works very well in practice \cite{Moldauer1968}.
This result was supported by Simonius \cite{Simonius1974} using Moldauer's sum rule \cite{Moldauer1967}.  The earliest formulations of Hauser-Feshbach theory which use the optical model began by noting that the energy averaged S-matrix $0\le \left|\left<S_{cc}\right>\right|\le 1$ so may be written written as \cite{Feshbach1954}
\begin{equation}
	\left<S_{cc}\right>=\exp{(-\eta_c+i\delta_c)},
\end{equation}
where $\eta_c$ is some positive, channel dependent, constant and $\delta_c$ is a channel dependent phase factor.
Implying 
\begin{equation}
	T_c=1-\exp{(-2\eta_c)}.
	\label{eq:originalTc}
\end{equation}
Expanding Eqs. \eqref{eq:opticalModelTc} and \eqref{eq:originalTc} in the weak coupling limit allows the identification of $x_c$ with $\eta_c$.

\subsection{Comparison of transmission coefficients}
% --------------------------------------------------
\label{subsec:compareTc}

The weak coupling limit ($x_c \ll 1$ and $T_c \ll 1$) can be achieved with very small widths or a small number of resonances widely spaced.  All three forms considered here (the SPRT, the Moldauer's Sum Rule and the Moldauer--Simonius forms) have the same first two terms in the weak coupling limit ($x_c\ll 1$):
\begin{equation}
	T_c^\textrm{weak}\approx 2x_c(1-x_c)
\label{eq:firstTwoTermsInWeakCouplingLimit}
\end{equation}
This is equivalent to equation (2.7) in the \emph{Atlas of Neutron Resonances} by S.F. Mughabghab~\cite{Atlas}.  
%NOTE: Moldauer and Simonius both use $\theta_c=2\xi_c$ and $\tau_c=2\eta_c$.

\begin{figure}
	\includegraphics[width=0.48\textwidth]{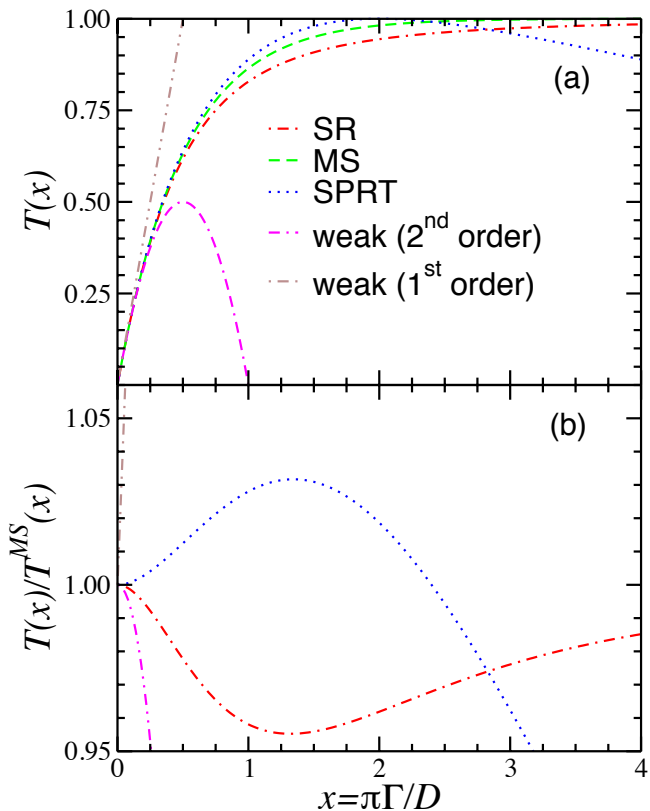}
	\caption{\label{fig:CompareTcFormulae2}Comparison the transmission coefficient formulae.  The top panel (a) shows the Moldauer--Simonius (MS), Sum Rule (SR), SPRT and linear ($1^{st}$ order) and quadratic ($2^{nd}$ order) approximations of the weak coupling transmission coefficients.  The bottom panel (b) shows the transmission coefficients in ratio to the Moldauer--Simonius (MS) transmission coefficient.}
\end{figure}

In Fig. \ref{fig:CompareTcFormulae2} we show the dependence of the transmission coefficient on $x_c$ for all three approaches as well as two different levels of approximation of the weak coupling limit.  In this figure, we see that if we keep only the leading order of the weak coupling expansion, we only match $T_c$ below $x_c\approx 0.25$.  The second order expansion of the weak coupling seems to work up to a much higher $x_c\approx 0.5$.   The other three forms are in roughly 5\% agreement up to $x_c\approx 2$ where the SPRT method diverges from the other two forms.

Given that $x_c$ is proportional to the ratio of the average channel width and the resonance spacing, the strong coupling limit ($T_c\rightarrow 1$) can be achieved either with very closely spaced resonances or with one or more resonances with anomalously large width(s).  
The transmission coefficients for both the Sum Rule and Moldauer-Simonius forms  approach 1 as $x_c\rightarrow \infty$ or equivalently $\overline{\Gamma}\gg D$ in the strong coupling limit.  
We note that the differences between the two parameterizations are at most 5\% deep in the overlapping resonance region ($\overline{\Gamma}_c\sim D$).

As we discussed above, the SPRT form predicts a maximum at $x_c=2$ and then decreases to 0 as $x_c\rightarrow \infty$.  The value of  $x_c=2$ corresponds to $\overline{\Gamma}_c\sim\frac{2}{3}D$, nearly in the overlapping resonance region.
In the SPRT approach, there is no explanation why the transmission coefficient should approach zero at large coupling other than that it is a natural outcome of the approximation in Eq. \eqref{eq:SthroughAveR}.

%Both the sum rule form (Eq. \eqref{eq:sumruleTc}) and the Moldauer-Simonius form give $T_c\rightarrow 1$ in the strong coupling limit.  
%Fig. \ref{fig:CompareTcFormulae2} provides a more detailed comparison of the two forms.  
%If there are many levels (as in the URR), in the MLBW approximation, $\tau_c\equiv\pi\left|\left<G_{cc}\right>\right|/D=\pi\overline{\Gamma}_c/D=\pi\rho(E)\overline{\Gamma}_c(E)$

Eqs. \eqref{eq:sumruleTc} and \eqref{eq:opticalModelTc} work in both strong and weak coupling limits.  Both provide us a method to directly compare the transmission coefficients used in a Hauser-Feshbach equation to the effective transmission coefficients in the RRR or URR.
% since the average resonance width can be computed directly from the resonances in an ENDF file's resolved resonance region or from the average resonance parameters in an ENDF file's unresolved resonance average parameters.   %We return to this feature in Section \ref{sec:application}.  

\subsection{Implications for the Hauser-Feshbach equation}
% --------------------------------------------------------
\label{sec:superradiance}

%Eqs. \eqref{eq:sumrulerhoc} or \eqref{eq:sumruleTc} build in the effects of the neutron scattering though an array of resonances.  %Therefore 
It is interesting to see how the different $T_c$ prescriptions modify Hauser Feshbach theory.  Both the Hauser Feshbach equation \eqref{eq:HF} and the WFC \eqref{eq:WFC} contain factors $\Gammabar_b/\sum_c \Gammabar_c$ which, by substituting \eqref{eq:sumruleTc} and using $x_c=\pi\Gammabar_c/D$, are modified to 
\begin{equation}
	\left.\frac{\Gammabar_b}{\sum_c \Gammabar_c}\right|_{\textrm{SR}}=\frac{T_{b}/\sqrt{1-T_b}}{\sum_{c}T_c/\sqrt{1-T_c}}.
	\label{eq:replacement}
\end{equation}
There is an additional factor of ${\Gammabar_a}$ in the absorption cross section which we will return to later in this section.  If instead one used the Moldauer-Simonius form of the transmission coefficient in Eq. \eqref{eq:opticalModelTc}, we arrive at a similar expression involving natural logarithms of $1-T_c$:  
\begin{equation}
	\left.\frac{\Gammabar_b}{\sum_c \Gammabar_c}\right|_{\textrm{MS}}=\frac{\ln{(1-T_{b})}}{\sum_{c}\ln{(1-T_{c})}}.
	\label{eq:replacement2}
\end{equation}
Both of these substitutions reduce to the one shown in Eq. \eqref{eq:replacementOld} in the weak coupling limit.  We note that the SPRT form does not provide a unique mapping between $x_c=\pi\overline{\Gamma}_c/D$ and $T_c$ due to its behavior at large $x_c$.  It is not clear how to make the same substitution that is done in Eqs. \eqref{eq:replacement} or \eqref{eq:replacement2} for the SPRT method.

With Eq. \eqref{eq:replacement}, the compound nuclear cross section becomes 
\begin{equation}
	\left.\sigma_{ab}^{\textrm{cn}}\rule{0cm}{0.5cm}\right|_{\textrm{SR}}=\sigma^{\textrm{abs}}_a\frac{T_{b}/\sqrt{1-T_b}}{\sum_{c}T_c/\sqrt{1-T_c}}{\cal W}_{ab}^{\textrm{SR}}(\vec{T}),
	\label{eq:HFmodSR}
\end{equation}
where ${\cal W}_{ab}^{\textrm{SR}}(\vec{T})={\cal W}_{ab}(\vec{\Gamma})$ denotes the usual WFC, using the replacement in equation \eqref{eq:replacement}.  
Similarly using Eq. \eqref{eq:replacement2}, the compound nuclear cross section becomes 
\begin{equation}
	\left.\sigma_{ab}^{\textrm{cn}}\rule{0cm}{0.5cm}\right|_{\textrm{MS}}=\sigma^{\textrm{abs}}_a\frac{\ln(1-T_b)}{\sum_{c}\ln(1-T_c)}{\cal W}_{ab}^{\textrm{MS}}(\vec{T}),
	\label{eq:HFmodMS}
\end{equation}
where ${\cal W}_{ab}^{\textrm{MS}}(\vec{T})={\cal W}_{ab}(\vec{\Gamma})$ denotes the usual WFC, using the replacement in equation \eqref{eq:replacement2}.  Either modification possibly has dramatic implications.  When we reach the strong coupling limit in only one channel (so $T_c\rightarrow 1$), that channel dominates the cross section.  The cross sections of all competing channels are strongly suppressed, an effect known as superradiance \cite{Auerbach2011}.  This effect may have been noted in a recent paper by Bertsch and Kawano \cite{Bertsch2018}, but they attributed it to a poor understanding of the lower bound on the number of fission exit channels.

The superradiant effect has been seen in many other mesoscopic systems and the question was raised in Ref.~\cite{Auerbach2011} why it is not seen in nuclear reactions.  We counter that it may well have been seen.  Since we treat compound nuclear reactions only in the weak coupling limit, so $T_c\approx 2\pi\Gammabar_c/D$, we have essentially neglected the functional dependence that gives rise to superradiance.  In the weak coupling limit we recover the traditional Hauser Feshbach equation with the WFC in Eq.~\eqref{eq:HF_Tc}.
%\begin{equation}
%	\sigma_{ab}^{\textrm{cn}}\approx\sigma^{\textrm{abs}}_a\frac{T_{b}}{\sum_{c}T_c}{\cal W}_{ab}
%\end{equation}
%This is equivalent to the treatment of the MC$^2$-II treatment for the average cross section in the URR.  We comment that the ENDF Format Manual's description of the MC$^2$-II method is valid only in the weak coupling limit, assuming SLBW resonances \cite{ENDFFormat}.

In practice, there are many effects that prevent $T_c\rightarrow~1$ and therefore mask superradiance.  We have already mentioned that direct reactions will lower the effective transmission coefficient.  Strong level repulsion keeps $D$ non-zero and therefore $\Gammabar_c/D$ finite.  Also, $\Gammabar_c\rightarrow\infty$ is unphysical.  However it can happen that a given optical model potential could lead to $T_c\approx 1$ for certain energies as we show in the next section.  This may be corrected by  refitting the optical model potential.  Given that one rarely uses the optical model in the URR, this region is understudied and the average cross section can easily be washed out by the cross section fluctuations.

It might be easier to see the effects of superradiance in the incoming channel simply because it is easier to control experimentally.  However, there are problems with this.
Blindly substituting the Sum Rule transmission coefficient \eqref{eq:sumruleTc} into the absorption cross section and using $x_a=\pi\Gammabar_a/D$, we have
\begin{equation}
	\left.\sigma^{\textrm{abs}}_a\rule{0cm}{0.5cm}\right|_{\textrm{SR}} = \frac{2\pi^2g_a}{k_a^2}\frac{T_a}{\sqrt{1-T_a}}
\end{equation}
which clearly is singular when $T_a\rightarrow 1$, violating both unitarity and common sense.  The Moldauer--Simonius transmission coefficient is not a viable alternative either as it gives
\begin{equation}
	\left.\sigma^{\textrm{abs}}_a\rule{0cm}{0.5cm}\right|_{\textrm{MS}} = \frac{2\pi^2g_a}{k_a^2}\left[-\ln(1-T_a)\right],
\end{equation}
which is also singular as $T_a\rightarrow 1$.
This issue was noted by Moldauer \cite{Moldauer1975b} and Englebrecht and Weidenm\"uller~\cite{Engelbrecht1973}.  In both cases, they  attributed it to the lack of ``M-cancellation'', that is, the effects of ignoring detailed level-level repulsion in the energy distribution of poles in the S-matrix.
Englebrecht-Weidenm\"uller transform does effectively reduce the transmission coefficient, preventing $T_c$ from reaching the singular value.  %Nevertheless, this makes us uncomfortable.  
One possible resolution may be that the entrance channels should not be treated symmetrically with the exit channels.   
Another potential resolution might require a detailed re-examination of how the WFC must be modified to account for level repulsion along the lines of Ericson {\em et al.}~\cite{Ericson2013,Ericson2016}.

% ===========================================================================
\section{Application to $^{90}$Zr}
% ===========================================================================
\label{sec:application90Zr}

We now turn to $^{90}$Zr, both because of its importance in nuclear energy applications and because of its simplicity.  $^{90}$Zr is nearly spherical, has a closed neutron shell, and has a relatively low level density.  In addition, the URR was recently re-evaluated by S.F. Mughabghab \cite{ENDFB80Library,Atlas}.  In our study, we computed the neutron transmission coefficients using the coupled channels code ECIS \cite{ECIS}, implemented in the EMPIRE code system \cite{EMPIRE}, and a Lane consistent dispersive soft rotor coupled channel optical model potential (RIPL OMP \#612) \cite{SoukhovitskiiUnpublished,RIPL}.

%The potential used in zr90 calculations is RIPL \#612, i.e., 
%\texttt{CC soft   Z=40  A= 90- 96    .0-200.0 MeV    E.Soukhovitskii}
%E.Soukhovitskii and R.Capote                                                    
%Unpublished.                                                                    
%New Lane consistent dispersive CC neutron OMP for incident energies between 1 keV 
%and 200 MeV for even Zr isotopes. Recommended for Zr-90-96. Local Fermi energy
% used. Soft rotor, OPTMAN should be used.

\begin{figure}
\includegraphics[width=0.48\textwidth]{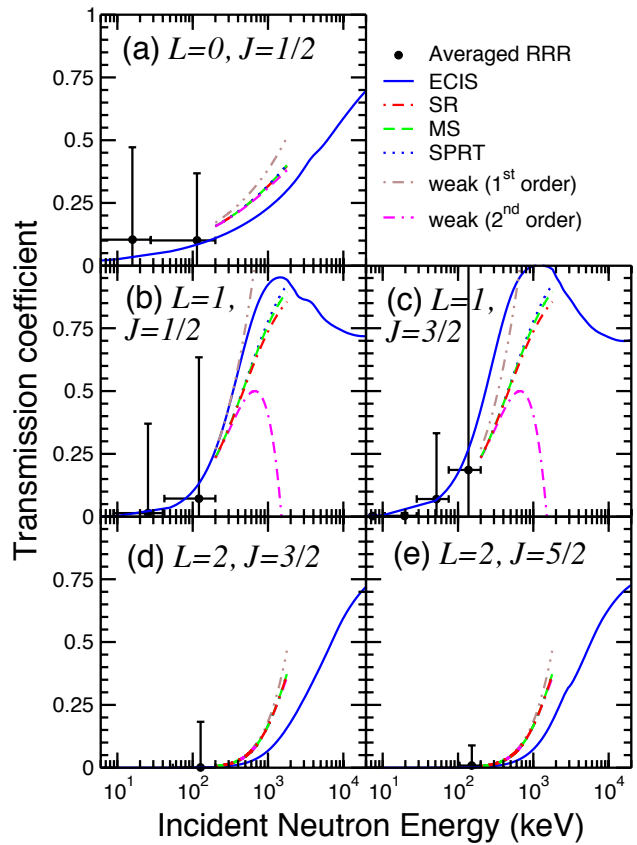}
\caption{\label{fig:nTcZr}Neutron transmission coefficients of $^{90}$Zr, computed using ECIS and the Lane consistent dispersive soft rotor coupled channel optical model potential (RIPL OMP \#612) \cite{SoukhovitskiiUnpublished,RIPL} and transmission coefficients computed directly from the resolved and unresolved resonance parameters in the ENDF/B-VIII.0 file.  The large uncertainty in the RRR is is expected due to the fact that the widths obey the Porter-Thomas distribution where the variance is $2\Gammabar$.}
\end{figure} 

In Fig. \ref{fig:nTcZr} we show the transmission coefficients extracted from the resolved and unresolved resonance parameters of the ENDF/B-VIII.0 $^{90}$Zr evaluation using the different transmission coefficient prescriptions discussed in this paper and using ECIS.  For $s-$, $p-$, and $d-$ wave neutrons impinging on the $0^+$ ground state of $^{90}$Zr, only the given $J$ shown in Fig. \ref{fig:nTcZr} are possible. 

The first aspect we note in Fig. \ref{fig:nTcZr} is that all of the transmission coefficient parameterizations are generally consistent at low energies but the two weak coupling approximations diverge from the rest of the forms above 500 keV.  The other three forms (SPRT, Moldauer-Simonius and Sum Rule) agree over the entire range of the URR and with the RRR at low energy.    In the RRR itself, the transmission coefficients are generally small, but with rather large variances.  This variance is a reflection of the intrinsic spread of the Porter-Thomas distribution of the resonance widths and not an artifact of limited statistics.  
%The average resonance parameters are given in Table \ref{table:90ZrRRR}.  
Finally, we note that the ENDF evaluation apparently uses a $J$ independent ratio of $\overline{\Gamma}_c/D$ even though $\overline{\Gamma}_c$ and $D$ both independently have a $J$ dependence.  %this is not documented clearly in the ENDF evaluation

\begin{figure*}
\includegraphics[scale=0.72,clip,trim=  0mm 11.5mm 4mm 0mm]{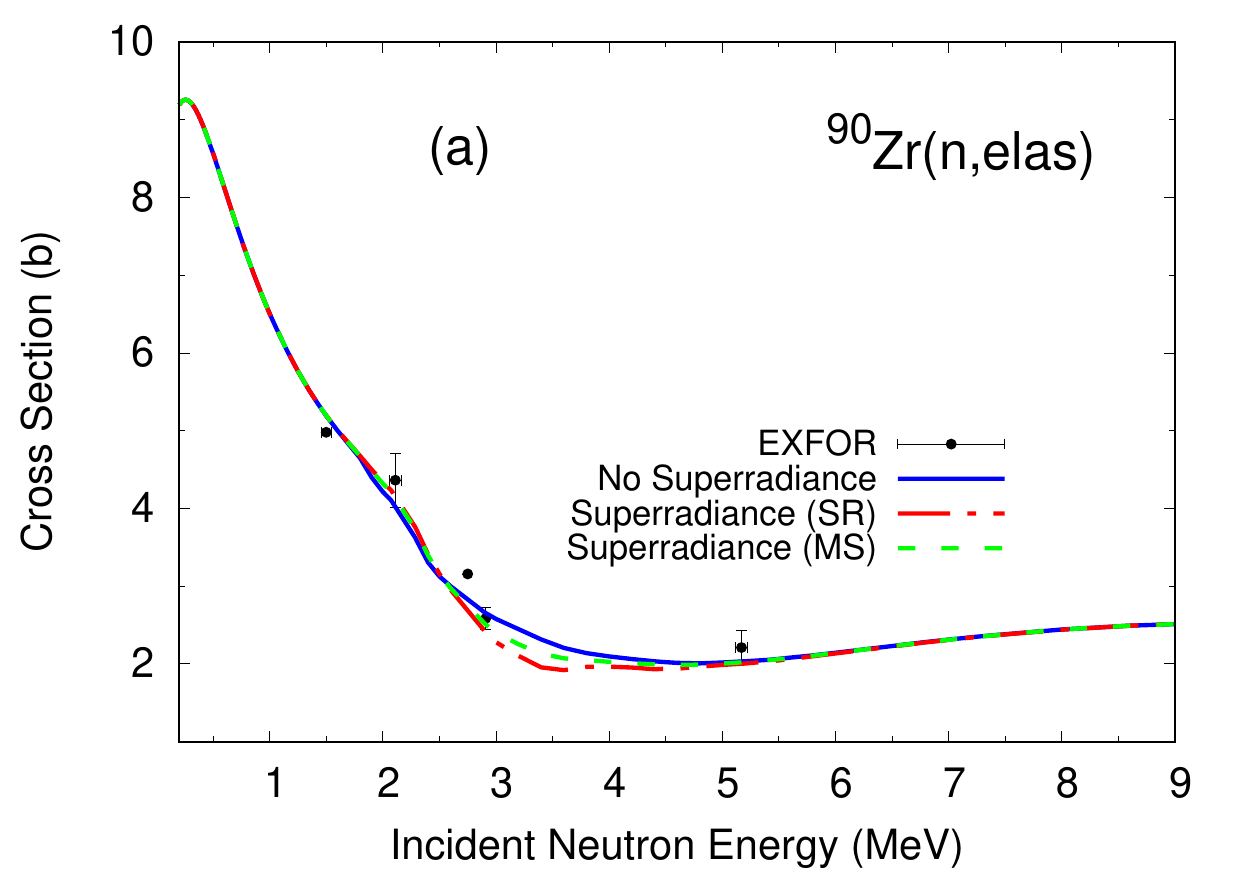}
\includegraphics[scale=0.72,clip,trim=  9mm 11.5mm 0mm 0mm]{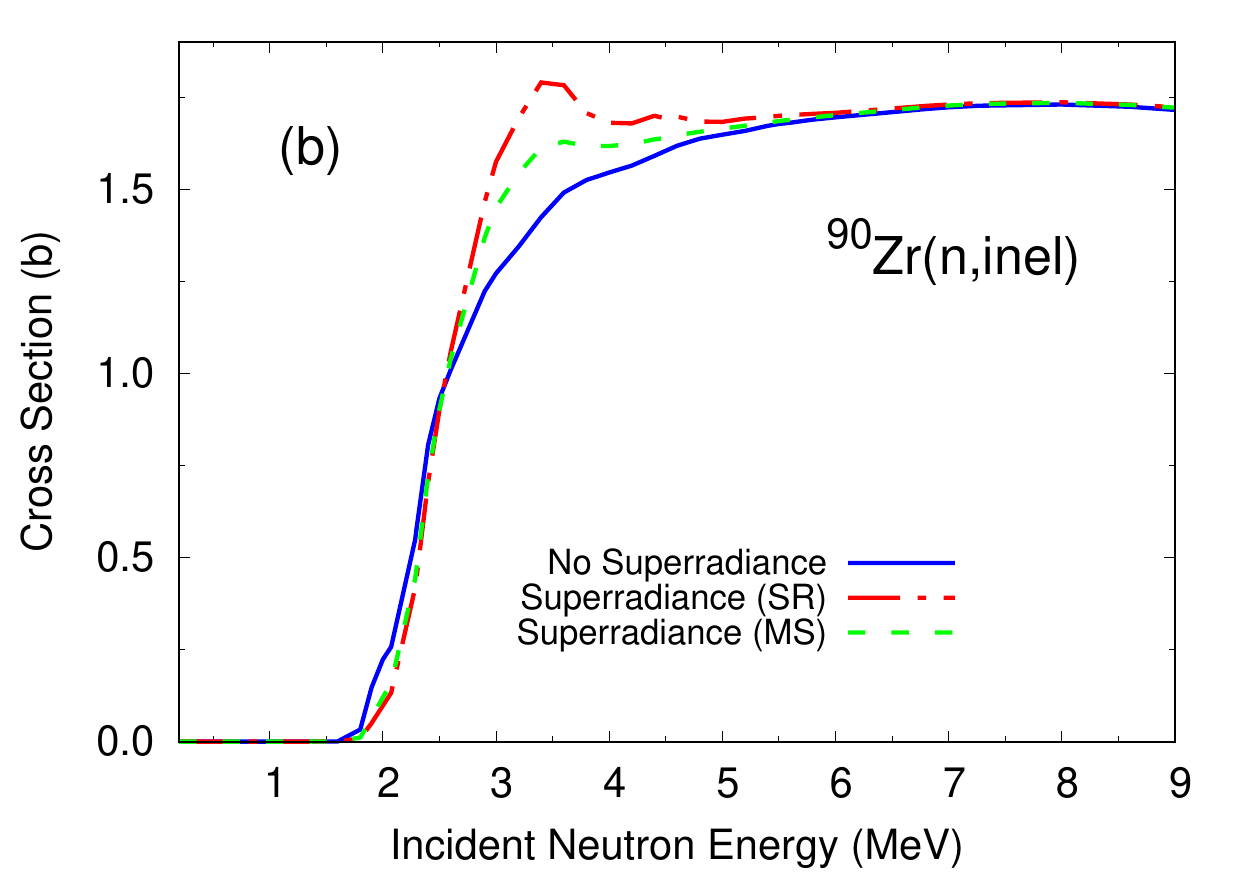}\\
\includegraphics[scale=0.72,clip,trim=  0mm    0mm 4mm 0mm]{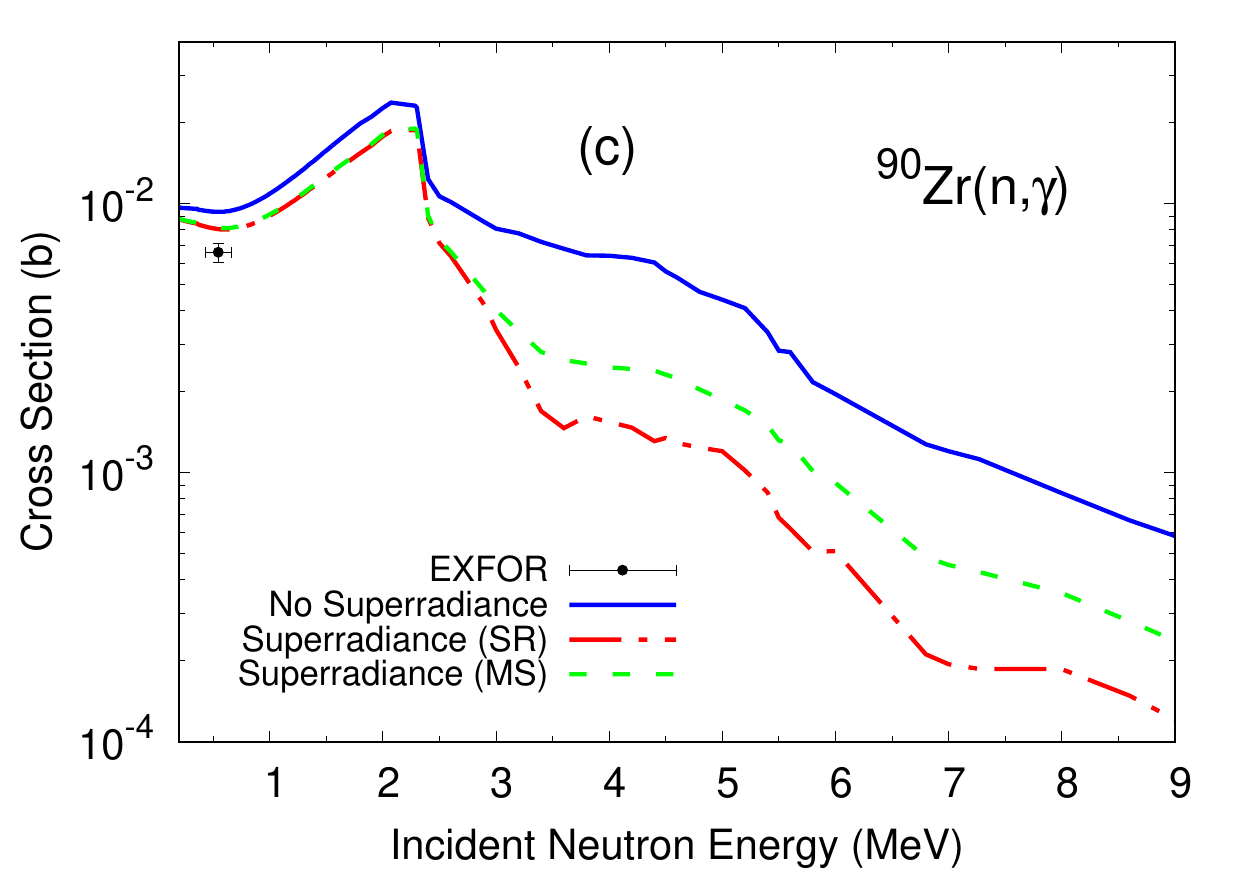}
\includegraphics[scale=0.72,clip,trim=  9mm    0mm 0mm 0mm]{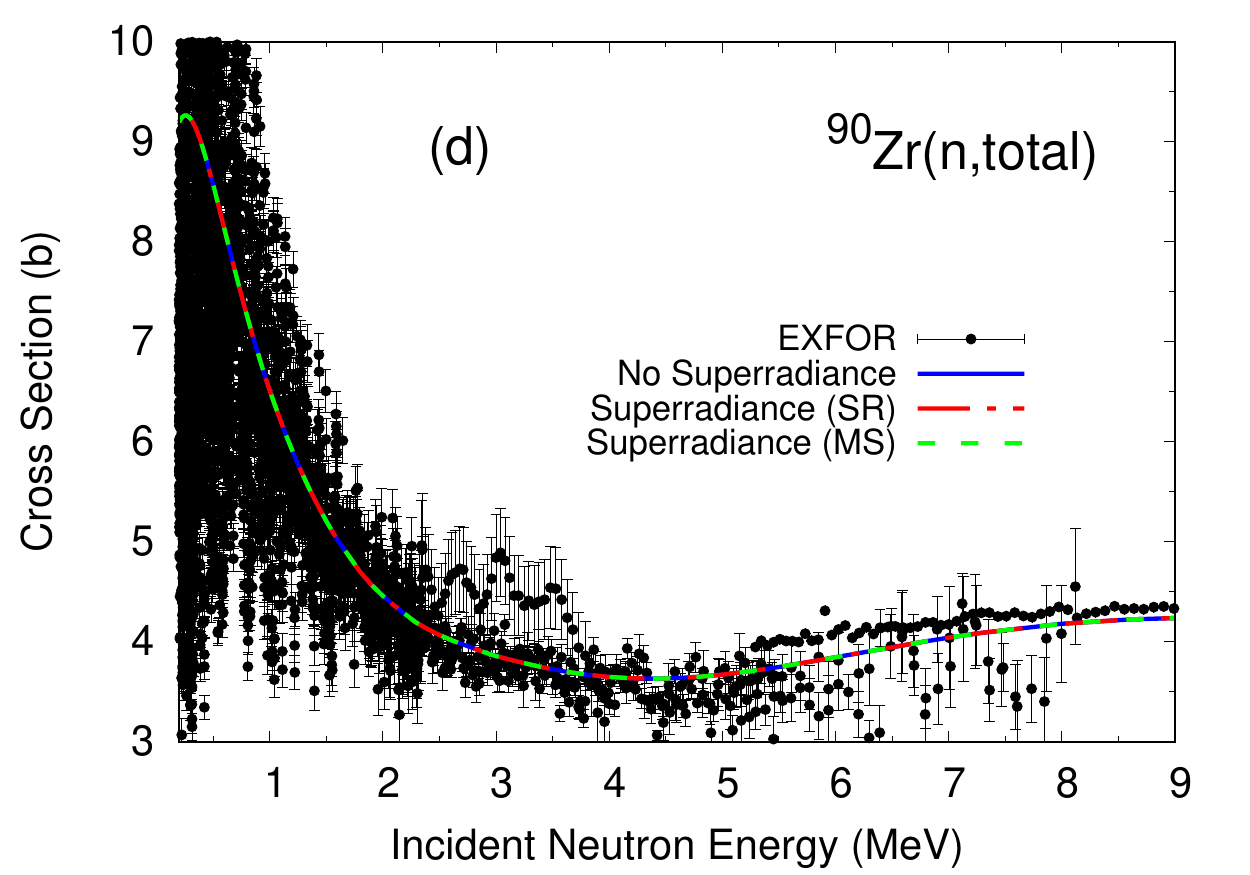}
\caption{\label{fig:ZrSigma} Plots of the (a) elastic, (b) total inelastic, (c) capture and (d) total cross sections, computed with and without the superradiance-modified Hauser Feshbach equation.  Superradiance cross sections computed with the Sum Rule form are labeled ``SR'' while those with the Modauer-Simonius form are labeled ``MS''.  Experimental data from the EXFOR library are also shown \cite{EXFOR}.}
\end{figure*} 

The transmission coefficients computed by ECIS clearly demonstrate a general consistency with the resolved and unresolved resonances, but disagree in detail.  The spin orbit coupling in the optical model potential does generate a $J$ dependence which is clearly visible in the plots, especially in the $p$-wave ($L=1$) channels.  We also note that the neutrons in Fig. \ref{fig:nTcZr} reach the strong coupling limit already at 1 MeV in the $p$-wave channels and could, in principle, exhibit superradiance.  We note that a coupled channels calculation with any realistic optical model potential should not let $T\rightarrow  1$ as this, when combined with all of the other channels in the problem, would violate unitarity.  In any event, this comparison provides a stringent test of the matching between the average resonance parameters and the optical model in the fast region.  

The peaks in the $p$-wave transmission coefficients are intriguing and suggest that we might find an indication of superradiance in the $^{90}$Zr cross sections.  We calculated the $^{90}$Zr cross sections using the EMPIRE \cite{EMPIRE} reaction code, using the Moldauer WFC, multi-step direct and multi-step compound reactions and Hartree-Fock Bogolyubov level densities from RIPL-3 \cite{RIPL}.  EMPIRE was modified to include the modified Hauser-Feshbach form in Eqs. \eqref{eq:HFmodSR} and \eqref{eq:HFmodMS} and the results are shown in Fig. \ref{fig:ZrSigma} for the total, elastic, capture and total inelastic cross sections.  In these plots, the effects of superradiance are not obvious either at low energy (where we are in the weak coupling limit) or at high energy (where there are a large number of open channels and the effects of pre-equilibrium emission become evident).  However, in the region around 2-4 MeV, we see noticeable differences between the cross sections computed with and without superradiance.  The most dramatic changes are in the total inelastic and capture cross sections, but the elastic cross section shows an effect as well.  In all cases, the difference to the non-superradiant cross section is greatest for the Sum Rule superradiant cross section.  We are not concerned about the change to the capture cross section as our calculations do not yet include the effects of semi-direct capture which will dominate over the compound contribution to the cross section at higher energies.  The total cross section, of course, shows no difference since unitarity must be preserved with or without superradiance.  However, the total cross section does indicate the size of the fluctuations which extend nearly to 5 MeV and, if measured, would be evident in the different partial cross sections.  

Superradiance appears to cause an interesting modification to the shape of the inelastic cross section just above threshold.   This shape has been seen in $(n,n'\gamma)$ measurements of other closed shell nuclei such as $^{56}$Fe (c.f. \cite{Negret2014}, Fig. 7), but is difficult to reproduce with traditional Hauser-Feshbach calculations.  It would be interesting to investigate the impact of superradiance in the CIELO Fe evaluation \cite{Herman2018}, however these cross sections are tightly constrained by other experimental data and would require a whole new evaluation.  In any event, a measurement of $^{90}$Zr$(n,n'\gamma)$ between 2 and 4 MeV would be very helpful by providing experimental evidence  (or lack of) of superradiance.
\section{Application to $^{197}$Au}
% ===========================================================================
\label{sec:application197Au}

The change to the capture cross section in Fig. \ref{fig:nTcZr}, although easily correctible in $^{90}$Zr, might have dramatic implications elsewhere.  Therefore we turn to $^{197}$Au where the capture cross section is regarded as a Neutron Data Standard \cite{Standards2018}.  

In what follows, we performed two sets of EMPIRE calculations, differing only in the choice of optical model potential.  We used the Delaroche dispersive coupled channel rigid rotor model (RIPL \#400) \cite{Delaroche1978} and the Koning-Delaroche potential \cite{KD}.  Both potentials give comparable results.  Also in these calculations, EMPIRE-specific level densities were used and the PCROSS\footnote{EMPIRE's PCROSS module includes the exciton model~\cite{Griffin1966}, based on the solution of the master equation \cite{Cline1971} in the form proposed by Cline \cite{Cline1972} and Ribansk\'{y} \cite{Ribansky1973}.} pre-equilibrium model was adopted for all outgoing particles.  The pre-equilibrium contribution was in general modest and only begins above 5 MeV incident energy.

\begin{figure}
	\includegraphics[width=0.48\textwidth]{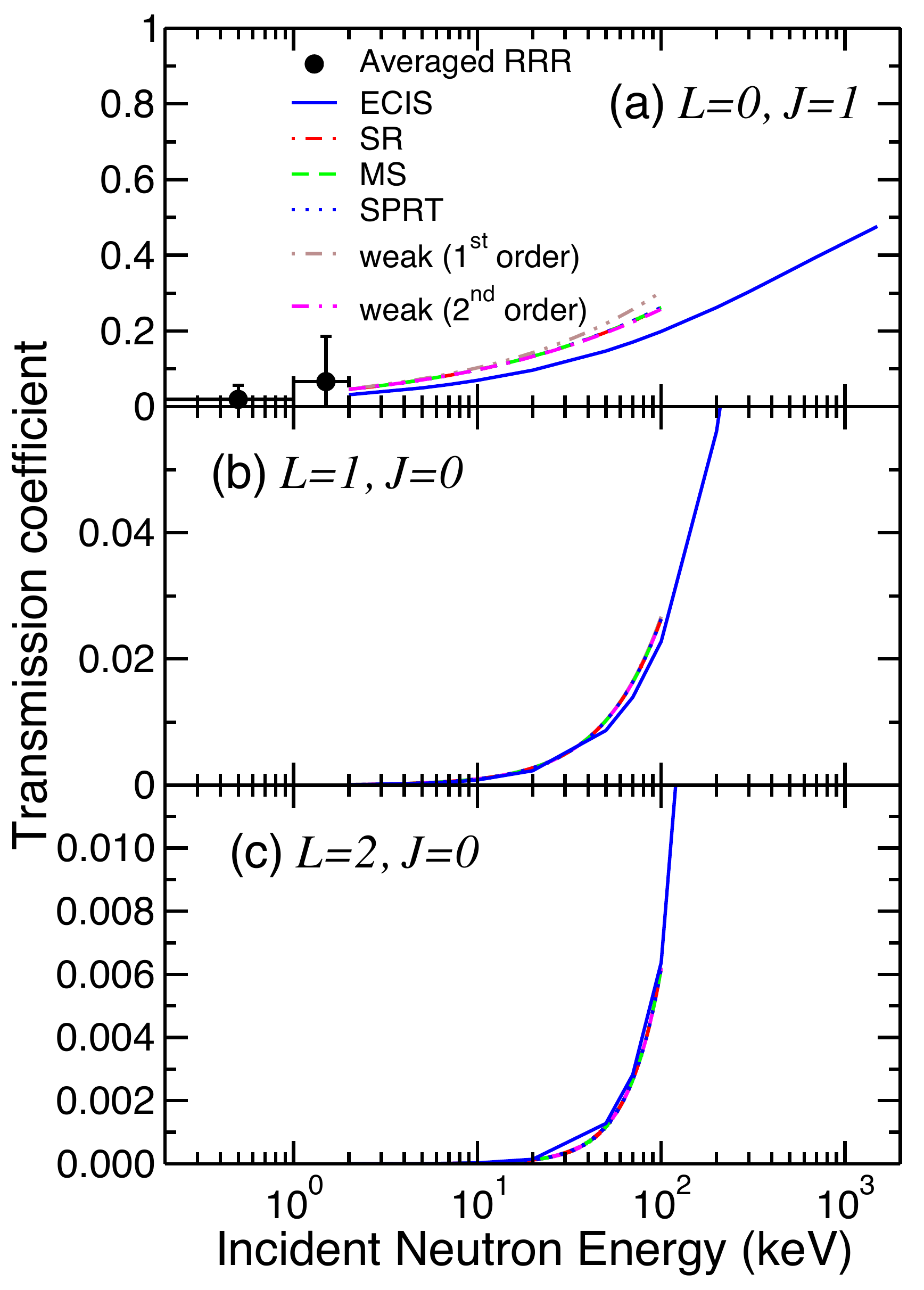}
	\caption{\label{fig:AuTc}Plots of the $J=0$ transmission coefficients extracted from the ENDF/B-VIII.0 $^{197}$Au evaluation's URR and RRR as well as that computed using ECIS and the \#400 optical model potential from RIPL-3 \cite{RIPL}.  Results from the Koning-Delaroche potential are equivalent.  As the neutron width is small for this nucleus for URR energies, all of our transmission coefficient prescriptions are also essentially equivalent and their curves are all superimposed on one another.}
\end{figure}

As in $^{90}$Zr, let us first examine the transmission coefficients before turning to the changes to the cross sections.  The $J=0$ transmission coefficients are shown in Fig. \ref{fig:AuTc}.  As in $^{90}$Zr, $\overline{\Gamma}_c/D$ was fixed for given $J$ in the $^{197}$Au evaluation although $\overline{\Gamma}_c$ and $D$ both independently have a $J$ dependence.  %his is not documented clearly in the ENDF evaluation.  Also is  a background cross section added, so comparison not fair.  background added to get standards cross section correct.  comparison is at best qualitative.  
At high energies (above the RRR), neutron scattering has weak or negligible dependence on the target nucleus spin.   Therefore, most optical model potentials treat the target nucleus as a $0^+$ nucleus and use $L$, $j$ and $\Pi$ as good quantum numbers (where $\vec{j}=\vec{L}+\vec{i}$ and $\vec{J}=\vec{j}+\vec{I}$).  If $I=0$, $\vec{J}=\vec{j}$.  However, the ground state of $^{197}$Au has $J^\Pi=3/2^+$.  Inside EMPIRE, additional weighting is done to couple up to the actual quantum numbers $\vec{J}=\vec{j}+\vec{I}$.  However, for the potentials under consideration, the transmission coefficient dependence on $j$, and hence $J$, was minimal over the energy range of the URR.  Therefore, only $J=0$ plots are shown.

In Fig \ref{fig:AuTc}, it is clear that there is overall good agreement between all of the transmission coefficients.  The largest deviations between the ECIS calculations and the URR transmission coefficients occurs in the $L=0$ panel, but even here the differences are modest.  The fact that the ECIS calculations reproduce the RRR and URR transmission coefficients is not really a surprise as Delaroche used the SPRT method to derive his optical model potential \cite{Delaroche1978}.

%\textcolor{red}{Statements about intrinsic spread in transmission coefficients from RRR hole here, but since the widths small, the variance is small}

\begin{figure*}
\includegraphics[scale=0.72,clip,trim=  0mm 11.5mm 4mm 0mm]{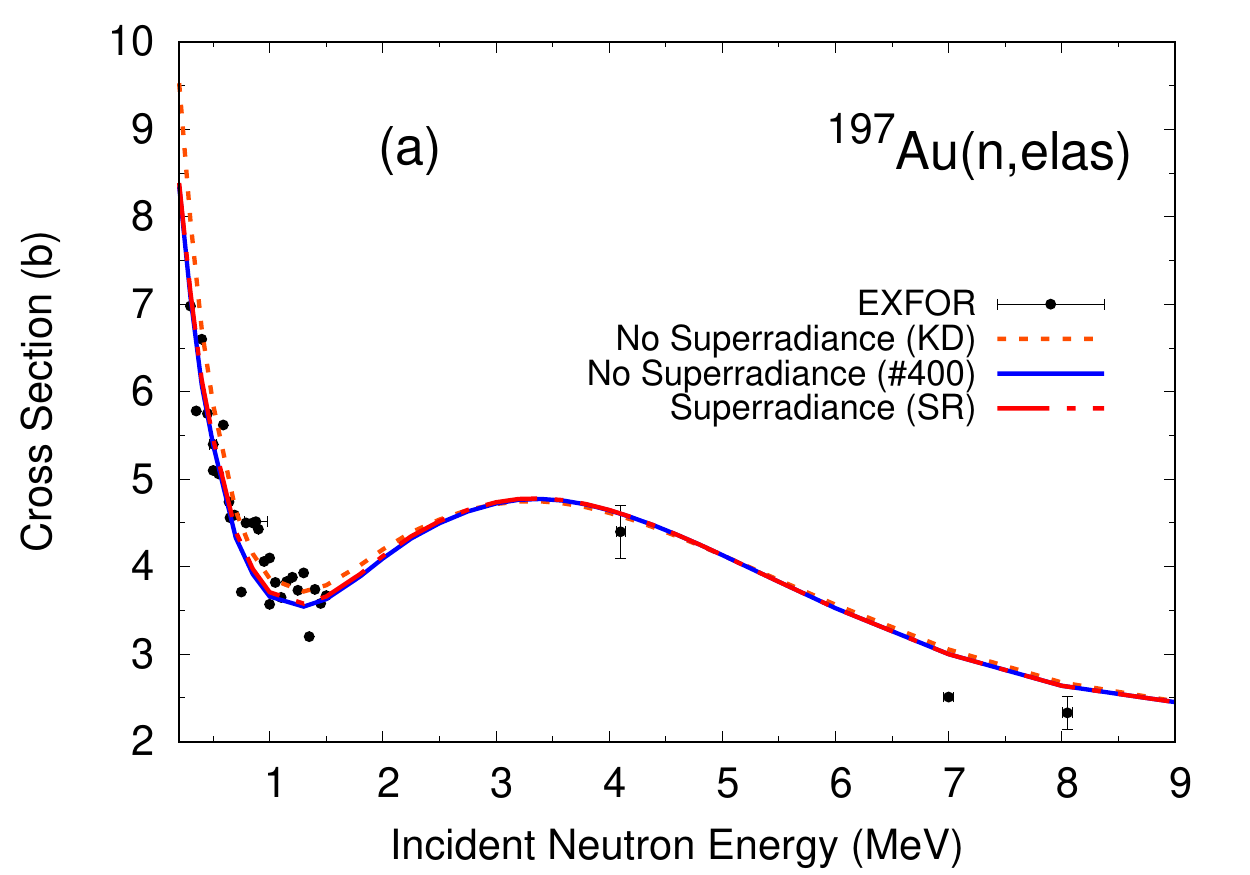}
\includegraphics[scale=0.72,clip,trim=  9mm 11.5mm 0mm 0mm]{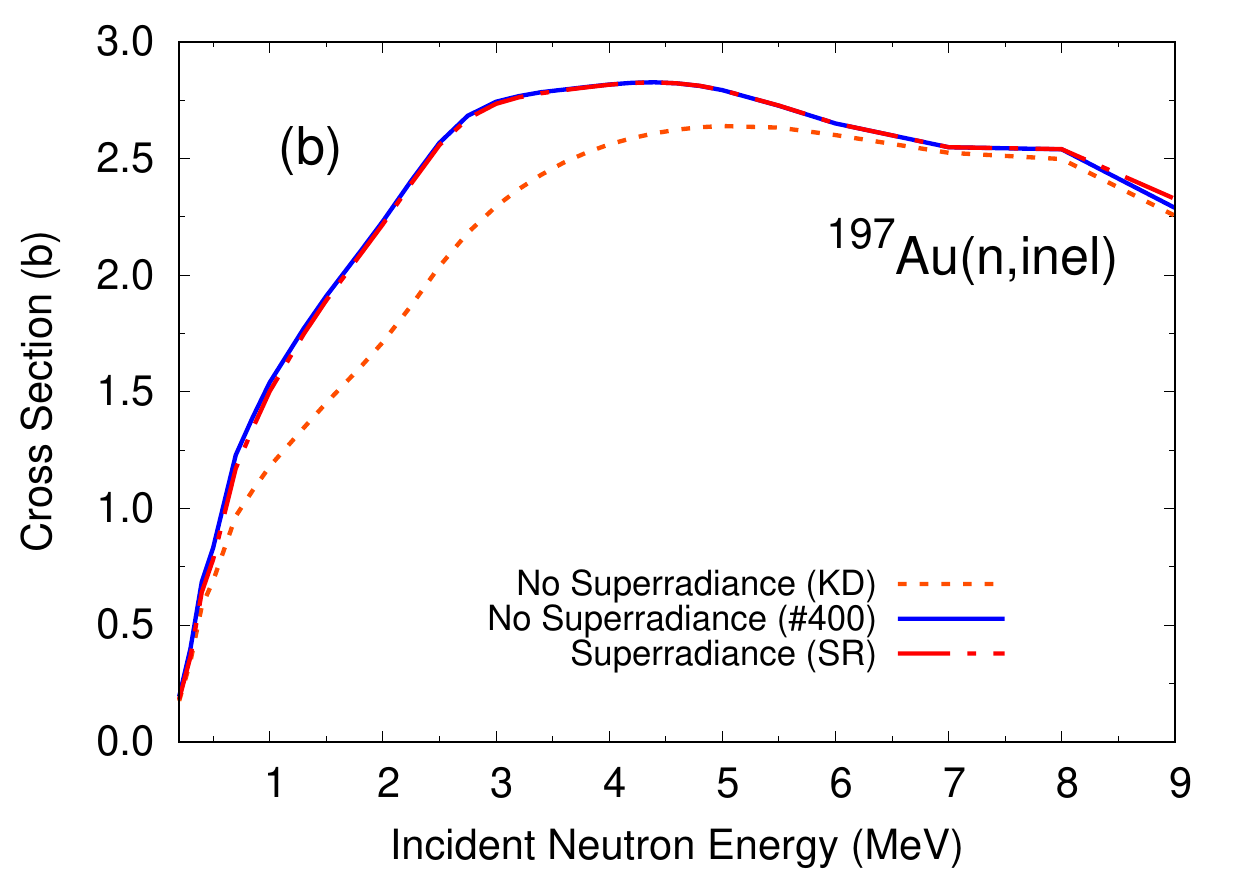}\\
\includegraphics[scale=0.72,clip,trim=  0mm    0mm 4mm 0mm]{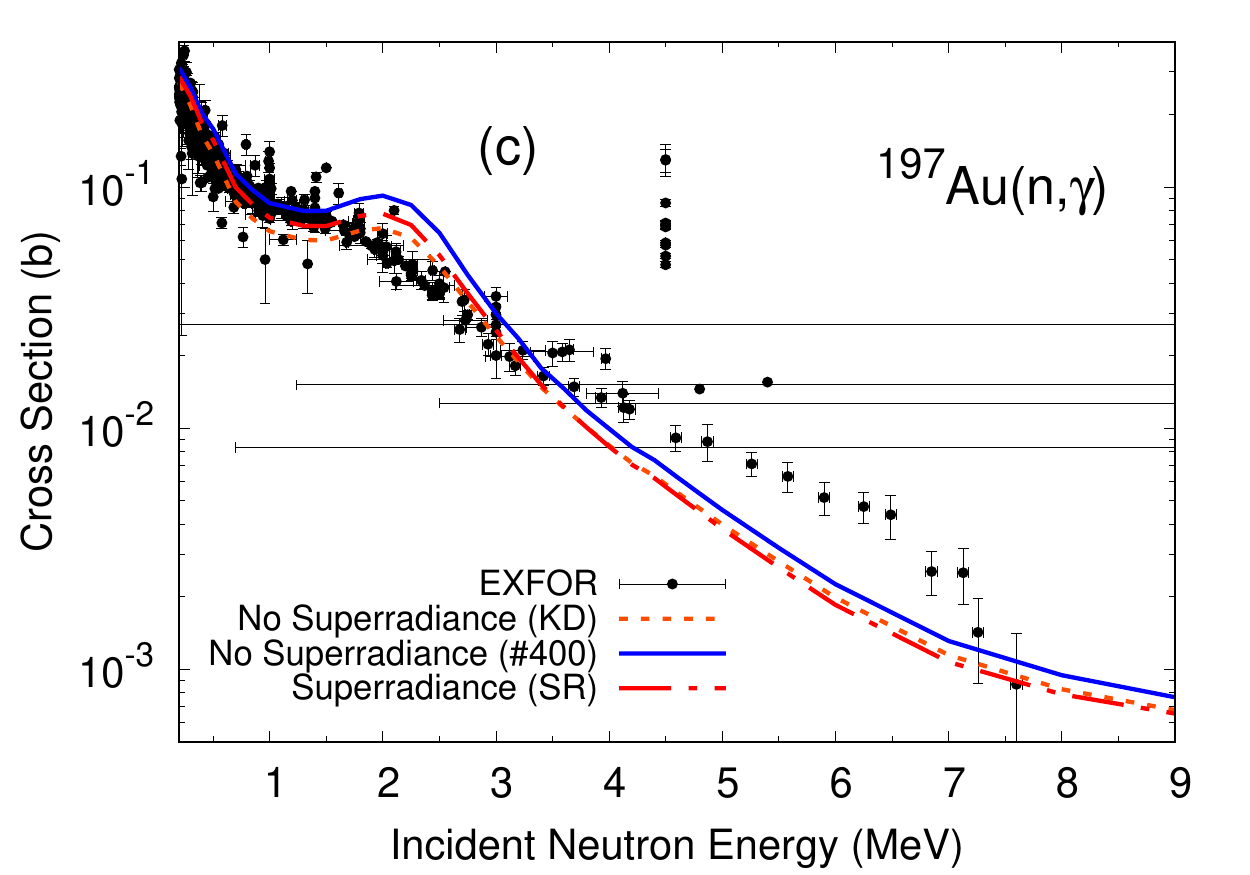}
\includegraphics[scale=0.72,clip,trim=  9mm    0mm 0mm 0mm]{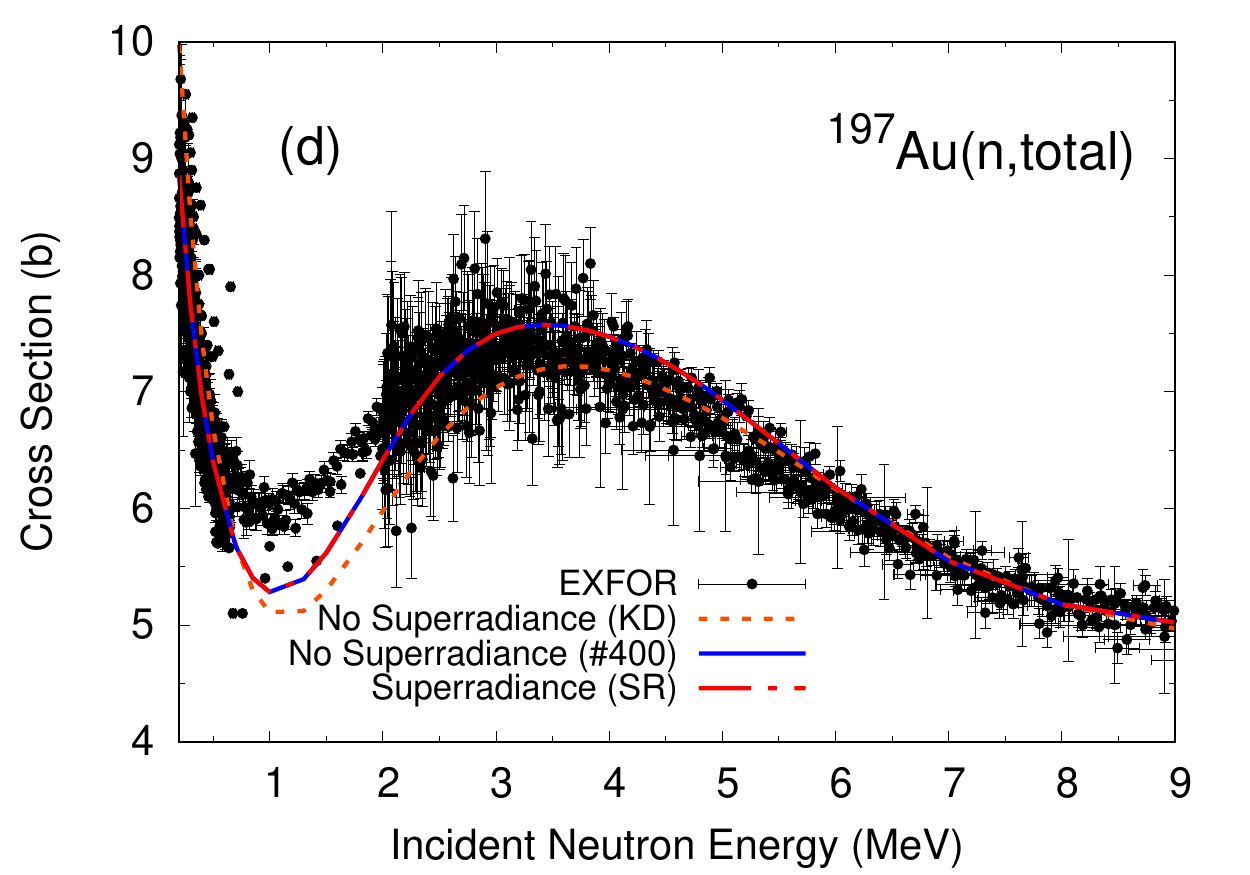}
\caption{\label{fig:AuSigma} Plots of the (a) elastic, (b) total inelastic, (c) capture and (d) total cross sections, computed with and without the superradiance-modified Hauser Feshbach equation.  In all panels, two non-superradiance calculations are shown: one using the \#400 optical model potential (labeled ``\#400'') and one using the Koning-Deleroche optical model potential (labeled ``KD'').  Experimental data from the EXFOR library are also shown \cite{EXFOR}.}
\end{figure*} 

In Fig. \ref{fig:AuSigma}, we show the $^{197}$Au elastic, total inelastic, capture and total cross sections both with and without the SUM Rule form of superradiance.  The difference between the Moldauer-Simonius and Sum Rule forms was not large enough to merit plotting both.  In the top two panels, it is clear the effect of superradiance on either the elastic or inelastic cross section is barely noticeable.  The effect on the capture cross section is larger as seen in the lower left panel.  We have no explanation for the bump at 2 MeV, but we do note that it is present both with and without superradiance and the difference between superradiant and regular results is roughly half the difference between results using different optical model potentials.  The panel on lower right shows that neither optical model potential describes the experimental data well between 500 keV and 2 MeV.  From an evaluators perspective, the differences in capture cross section can be easily corrected by modifying either the gamma-ray strength function or the level densities or by adding background as was done in the ENDF/B-VIII.0 evaluation.

The difference in impact of superradiance on $^{90}$Zr and $^{197}$Au is easy to explain.  $^{90}$Zr has very few open neutron channels and large $p$-wave neutron transmission coefficients near 1 MeV.  On the other hand $^{197}$Au has 11 open neutron channels with $J\le 2$ and all are of comparable magnitude.

%\begin{verbatim}
%197Au
%Following options/parameters have been used
% -------------------------------------------
% Main calculations output control set to  3
% Number of energy steps in the integration set to  80
% ENDF formatting disabled
% Recoils are not calculated
% EMPIRE-specific level densities (J>>K aprox.) selected 
% Moldauer width fluctuation correction with Kawano-Talou (NDS118, 183, 2014) nu selected
% Tlj coupling for the top CN bin up to incident  9.00 MeV
% Coupled Channels Method used for Tl calcul. in outgoing ch.
% Optical model parameters for direct inelastic scattering set  to RIPL # 400
% Exciton model calculations with code PCROSS
% Cluster emission in terms of the Iwamoto-Harada model
% Kalbach systematics angular distributions (see RIPL-1)
% Mean free path parameter in PCROSS set to  1.5 (Recomm:~1.5)
% CN decay and Direct cross sections added incoherently
% CN anisotropy calculated using Blatt-Biedenharn coefficients
% Traditional Hauser-Feschbach Tlj_{CC}(E)=Tlj_{GS}(E-Egs)
%\end{verbatim}

% ===========================================================================
\section{Conclusion}
% ===========================================================================
\label{sec:conclusion}

We have investigated the consequences of different formulations of the neutron transmission coefficient, enabling a rigorous connection between the resolved resonance, unresolved resonance and fast regions for neutron-induced reactions.  Our work shows that if one has predictions for the mean level spacing (or, equivalently, level density) and an optical model potential, then one can predict the average neutron widths using Eqs. \eqref{eq:sumruleTc} or \eqref{eq:opticalModelTc}.  This provides a tool for predicting neutron widths far off stability.
%, with significant possible astrophysical impacts.   
This work also sheds a light on the large coupling behavior of the SPRT method and why it should be avoided in strong coupling cases.

Our work also shows how and where superradiance may impact nuclear reactions.  Features of superradiance may be present in both the elastic and inelastic channels but are obscured for a variety of reasons.  First and foremost, it is experimentally difficult to disentangle elastic and inelastic neutrons and only the recent experiments measuring $(n,n'\gamma)$ may be able to see the signatures of superradiance.  Furthermore, while we see superradiance in our calculations on $^{90}$Zr, the large  fluctuations in the cross section undoubtably obscure it.  In $^{197}$Au, superradiant effects are even smaller than in $^{90}$Zr and can even be corrected away by an evaluator to allow a match to experimental data.  The effects of superradiance in the compound nuclear cross section appears to be small in most cases and is only evident in systems with a small number of open channels with large transmission coefficients.

Several questions remain as to the role of correlations in the level spacing distribution and the width fluctuation factor.  Answering these questions will likely help us understand the correct behavior of the absorption correction.  This might also help us understand the variance of the cross section in the URR.  Finally, we would like to extend our work to other types of channels, especially capture and fission.

% ===========================================================================
\section*{Acknowledgements}
% ===========================================================================

The authors wish to acknowledge the fruitful conversations with E. David Davis and Toshihiko Kawano.
Work at Brookhaven National Laboratory was sponsored by the Office of Nuclear Physics, Office of Science of the U.S. Department of Energy under Contract No. DE-AC02- 98CH10886 with Brookhaven Science Associates, LLC.

% ===========================================================================
\appendix
\section{Appendix}
% ===========================================================================

In the main text, we consider two sums over the resonance widths which we approximate as ensemble averages:
\begin{eqnarray}
	\frac{\Delta E}{D}\left<\left<\Gamma_c\right>\right>&\approx&\sum_\lambda \Gamma_{\lambda c} = 2P_c \sum_\lambda \gamma_{\lambda c}^2,\label{aeq:1}\\
	\frac{\Delta E}{D}\left<\left<\Gamma_c^{1/2}\Gamma_{c'}^{1/2}\right>\right>&\approx&\sum_\lambda \Gamma_{\lambda c}^{1/2}\Gamma_{\lambda c'}^{1/2}.\label{aeq:2}
\end{eqnarray}
Here the ensemble average of $f(X)$ is just the expectation value of $f(X)$ using the probability distribution function for $X$, namely  
$\left<\left<f(X)\right>\right>=\int dX{\cal P}(X) f(X)$.  

We assume the widths are distributed according to the generalized Porter-Thomas distribution, a.k.a. a $\chi^2$ distribution with $\nu$ degrees of freedom
\begin{equation}
	{\cal P}^{PT}(\Gamma_c|\overline{\Gamma}_c, \nu) d\Gamma_c = \frac{e^{-y}y^{\nu/2}}{\Gamma(\nu/2)}\frac{dy}{y}
\end{equation}
where $y=\frac{\nu}{2}\frac{\Gamma_c}{\overline{\Gamma}_c}$.  Here, $\Gamma(z)$ is a Gamma function and should not be confused with the widths $\Gamma_c$.
With this distribution, it is straightforward to show that 
\begin{equation}
	\left<\left<\Gamma_c^n\right>\right>=\left(\frac{2\overline{\Gamma}_c}{\nu}\right)^n\frac{\Gamma\left(\nu/2+n\right)}{\Gamma\left(\nu/2\right)}.\label{aeq:3}
\end{equation}

We now turn to Eq. \eqref{aeq:1}.  Specializing to $n=1$, we find $\left<\left<\Gamma_c\right>\right>=\overline{\Gamma}_c$ and 
\begin{equation}
	\sum_\lambda \Gamma_{\lambda c} = 2P_c \sum_\lambda \gamma_{\lambda c}^2\approx\frac{\Delta E}{D}\overline{\Gamma}_c.\label{aeq:4}
\end{equation}
This proves Eq. \eqref{eq:sc} in the main text.

Eq. \eqref{aeq:2} is not as easy to tackle.  While we do know the probability distribution for a single width, the joint probability of two widths ${\cal P}(\Gamma_c, \Gamma_{c'})$ is unknown.  At best, we can assume that the widths of different channels are uncorrelated (an approximation that is untrue as the main text argues).  With this incorrect assumption, ${\cal P}(\Gamma_c, \Gamma_{c'})\approx{\cal P}^{PT}(\Gamma_c){\cal P}^{PT}(\Gamma_{c'})$ and
\begin{equation}
	\left<\left<\Gamma_c^{1/2}\Gamma_{c'}^{1/2}\right>\right>\approx\left<\left<\Gamma_c^{1/2}\right>\right>\left<\left<\Gamma_{c'}^{1/2}\right>\right>.
\end{equation}
With this, we find 
\begin{equation}
\begin{split}
\lefteqn{\left<\left<\Gamma_c^{1/2}\Gamma_{c'}^{1/2}\right>\right> \approx}&\\
	& \;\;\;\;\;\left(\frac{4\overline{\Gamma}_c\overline{\Gamma}_{c'}}{\nu_c\nu_{c'}}\right)^{1/2}
	\frac{\Gamma((\nu_c+1)/2)\Gamma((\nu_{c'}+1)/2)}{\Gamma(\nu_c/2)\Gamma(\nu_{c'}/2)}.
\end{split}
\end{equation}

% ===========================================================================
% Bibliography
% ===========================================================================

\end{document}